\documentclass[12pt,preprint]{aastex}
\usepackage[left=1in,top=1in,bottom=1in,right=1in,nohead,nofoot]{geometry}
\usepackage{multicol}

\begin{document}

\begin{center}{\bf Detecting and Characterizing Planetary Systems with Transit Timing}\\
Jason H. Steffen, B. Scott Gaudi, Eric B. Ford, Eric Agol, Mathew J. Holman
\end{center}

{\bf 1. Abstract}

In the coming decades, research in extrasolar planets aims to
advance two goals: 1) detecting and characterizing low-mass planets
increasingly similar to the Earth, and 2) improving our understanding
of planet formation.  We present a new planet detection method that is capable of making large advances towards both of these objectives and describe a modest network of telescopes that is able to make the requisite observations.  In a system where a known planet transits its host star, a second planet in that system will cause the time between transits to vary.  These transit timing variations can be used to infer the orbital elements and mass of the perturbing planet even if it has a mass that is smaller than the mass of the Earth.  This detection technique complements other techniques because it is most sensitive in mean-motion resonances where, due to degeneracies, other techniques have reduced sensitivity.  Small ground-based observatories have already exceeded the photometric precision necessary to detect sub-Earth mass planets.  However, TTV planet searches are currently
limited by the relatively small number of high-precision transit data and insufficient observing time on existing telescopes.  These issues will be compounded as the number of known transiting
planets suitable for TTV study will increase substantially in the near future.  A relatively modest
investment in a ground-based network of small ($\sim 0.5~{\rm m}$
telescopes could provide the needed coverage and so dramatically
increase the effectiveness of transit timing observations.

{\bf 2. Introduction: Planetary Transits}


After many years of effort, the transit surveys for planets have recently come
to fruition (see Charbonneau et~al.\ 2007 for a review).  To date 9 transiting planets 
orbiting bright ($V\la 12$) stars are known, most of these were discovered through these transiting
planet surveys (Alonso et al.\ 2004, McCullough et al.\ 2006, O'Donovan et al.\ 2006, Bakos et al.\ 2007, Collier-Cameron et al.\ 2007). Beyond the existing ground based transit surveys, the recent
launch of ESA's CoRoT satellite (Baglin et~al.\ 2002) and the planned
launch of NASA's Kepler mission (Borucki et~al.\ 2003) are expected to
discover many additional transiting systems.

Once a star has been discovered to host a transiting planet, there is
the potential for many exciting follow-up observations (see Charbonneau et~al.\ 2007).  In this white paper,
we focus on the potential for transit timing observations to search
for additional planets.  If a star harbors a single transiting planet,
then the transit times would be strictly periodic.  On the other hand,
if a star harbors more than one planet, then the planets' mutual
gravitational perturbations will cause their orbits to deviate from
simple Keplerian orbits and hence from strictly periodic times of transit. These transit
timing variations (TTV) can be used to infer the orbital elements of
the perturbing planet (Agol et al.\ 2005, Holman \& Murray 2005).  Since 
very small changes in a planet's orbital elements can be measured by
timing transits, this represents a very powerful method for
searching for low-mass planets.  The sensitivity of the TTV method is
further enhanced if the perturbing planet is near a mean-motion resonance (MMR) where the timing variation depends on the planet-planet mass 
ratio rather than planet-star mass ratio.  For example, for a transiting, 
Jupiter mass planet on a 3-day orbit, an Earth mass planet in the 2:1 
resonance will cause periodic variations in the transit times that have 
an amplitude greater than one minute (See Figure 1).  This should be compared to 
the (best) transit timing precision of
$\sim$10s that has been demonstrated using small (1.2m) ground-based
observations (Holman et al.\ 2006).  Thus, the TTV technique is already
capable of probing for planets with masses even less than the mass of
the Earth (Agol \& Steffen 2007).  Further, the enhanced sensitivity
of the TTV method resonant planets is particularly
interesting to theorists, since other dynamical detection techniques
often have a reduced sensitivity to such planets due to potential
degeneracies between orbital parameters.

The TTV technique was first applied to transit measurements of the
TrES-1 planetary system.  In this case, a set of 11 ground-based
observations were able to probe for terrestrial mass planets near
several interesting MMR's (Steffen \& Agol 2005).  In a similar
study, 13 Hubble Space Telescope observations of the HD 209458 system
were sensitive to planets with masses approaching that of Mars (Agol \& Steffen 2007).
We note that the ground based observations (using KeplerCAM on
the 1.2m Fred Whipple Observatory) of XO-1 (Holman et al.\ 2006) are of higher
precision than some of the HST observations of the HD~209458
system---indicating the great potential for high quality transit
timing observations, even with modest ground-based telescopes.
The highest transit-timing precision to date, $\sim$ 6 seconds, has been 
recently obtained for HD 189733b with the {\it Spitzer} Space Telescope
(Knutson et al.\ 2007);  however, all other known transiting planets would 
have a lower timing precision due to the faintness 
of their host stars in the {\it Spitzer} bandpass.


{\bf 3. Scientific Justification for TTV Observations}

Analyzing the variations of transit times in planetary systems can
address many important questions regarding planetary populations,
characteristics, planet formation, and evolution theories.  We address
several of these questions here.

{\bf Planet Detection:}
 
$\bullet$ The TTV method can increase the scientific value of
other transit search programs such as NASA's {\it Kepler} mission. It provides
the capability to discover additional planets around stars with a
transiting planet, even if the second planet does not transit the
star.  This could be particularly useful for detecting
terrestrial-mass planets in the habitable zone where such a planet has a
reduced geometric probability of transiting.  This is accomplished by
analyzing the transits of a planet with an orbit that lies interior to
the habitable zone, where planets are more likely to
be transiting.  For a conservative example, if a Jupiter mass planet
were on a 130-day orbit, high quality ground-based observations of the
transit times could identify a terrestrial mass perturbing planet on a
non-resonant, low eccentricity orbit ($\sim 0.05$) at 1 AU (see Figure 1).  From this
conservative starting point, the noise remains constant while the TTV
signal scales linearly with the mass of the perturber, the
eccentricity of the perturber, and the period of the transiting
planet.  The signal also grows sharply as the system approaches one of
several MMR's.

$\bullet$ Transit timing can be
quite sensitive to low-mass planets.  Thus, the TTV method can test
theoretical models of planet formation and migration predict that
terrestrial mass planets should be common around stars with a
short-period giant planet (e.g., Narayan et al.\ 2005).

$\bullet$ Although it is possible to 
search for additional planets in each system
by directly observing their transits (e.g., Brown et al 2001),
this generally requires continuous, precise
photometry {\it outside} of transit, and so a
substantial commitment of observation time.  Since the TTV method only requires photometric
observations near the times of transit, it provides an efficient tool
to search for additional companions.  

{\bf Planet Formation Theory:} 

$\bullet$ Transit timing observations are particularly
valuable because they are extremely sensitive to planets near
MMR.  The core accretion
model of planet formation predicts that terrestrial mass planets will
often be caught in MMR with a gas giant planet as
that planet migrates inward (Zhou et~al.\ 2005, Thommes 2005, Cresswell \& Nelson 2006, Terquem \& Papaloizou 2007,  Fogg \& Nelson 2007, Mandell
et al.\ 2007).  This occurs
either by: 1) the migrating planet sweeping smaller planets into
interior resonances or 2) exterior planets interacting with the gas
disk and rapidly spiraling toward the giant and becoming trapped in
exterior resonances.  Thus, if the core-accretion model is correct, we
expect that many stars with transiting planets may also have
lower-mass planets in interior and/or exterior resonances.  Further,
the details of the trapping in different resonances depend on the
migration timescale, the disk lifetime, and the eccentricities and inclinations of both
bodies at the time the bodies approach resonance.  Thus, studying the
frequency of planets in various resonances can provide constraints on
the migration process.  (This is similar to the constraints on the
timescale and extent of migration for Neptune based on the resonant
Kuiper belt objects).  

$\bullet$ Provided that sufficient TTV observations are
made, a null result would also be significant.  While
the core-accretion model predicts that trapped terrestrial planets
should be common, the gravitational instability
model of planet formation does not anticipate the presence of
such planets in any quantity.
Thus, the discovery of small objects in MMR with
transiting planets---precisely the regime where the TTV technique is
most sensitive---would support the former theory (Zhou et~al.\ 2005).  The lack of such
planets would support the latter or place strong constraints on
parameters of migration theory, particularly if the period of the
transiting planet were a few tens of days (tidal effects may affect
systems with shorter orbits which we discuss next).  

$\bullet$ The TTV method could measure the ubiquity of closely packed 
planetary systems (e.g.\ Juric \& Tremaine 2007) and/or study the 
dynamical properties of systems with strongly interacting planets. 
Small planets on interior resonant
orbits with a hot Jupiter (or the lack thereof) would constrain models
of tidal interactions that may cause the orbit of the inner planet to
decay (D.\ Fabrycky, private communication).  

{\bf Characterization of Planetary Systems:}  

$\bullet$ From TTV analyses one may be
able to identify the mutual inclination of planetary systems 
(Miralda-Escud{\' e} 2002).  That
quantity has implications for allowed mechanisms for the growth of
eccentricities in planetary systems (e.g., Chatterjee
et al.\ 2007).  It can also provide a determination of the mass of a
non-transiting perturbing planet, something that is very difficult to
identify with other planet detection techniques.  

$\bullet$ For systems where
multiple planets transit there can be sufficient information to determine
the absolute masses and radii of the two planets and the host star independent 
of stellar models, as with double-lined eclipsing binaries.
This is crucial for determining when terrestrial-sized planets have 
terrestrial-mass (which would otherwise be challenging or impossible with 
the radial-velocity technique) and allows for a measurement of their densities.

$\bullet$ Knowledge of the mass and
orbit of additional planets can help interpret observations of the
transiting planet.  For example, it has been proposed that planets
with unexpectedly large radii and being heated by a combination of
tidal dissipation in the planet and orbital perturbations from other
planets (e.g., Bodenheimer et al.\ 2001).

{\bf 4. Requirements for a Successful Transit Timing Program}

{\bf Transit Observations:}

Searching for planets with the TTV method requires precise
photometric observations during many transits.  For
a photometric precision $\sigma_{ph}$, the central time $T_c$ of a given 
transit can be measured with an accuracy (Ford \& Gaudi 2006, Holman
\& Murray 2005):
$\sigma_{T_c} \simeq [t_e/(2\Gamma)]^{1/2}\sigma_{ph}\rho^{-2},$
where $t_e$ is the duration of ingress/egress, $\Gamma$ is the rate
at which observations are taken,
and $\rho$ is the ratio of the planet radius to stellar
radius (this equation ignores limb-darkening).  For typical parameters and millimagnitude photometry ($\sigma_{ph} \sim 10^{-3}$), 
$T_c$ can be measured to tens of seconds (e.g., Brown et al.\ 2001; Holman et al.\ 2006). 
Such precision enables the detection of Earth-mass companions to transiting 3-day period gas giants for optimal (i.e.\
resonant) configurations (Steffen \& Agol 2005).  For longer period transiting planets, the sensitivity is correspondingly
better, scaling as $m_2 \propto P_1^{-1}$, where $1,2$ label the transiting
and perturbing planets.

Achieving millimagnitude photometry on stars of $V\sim 11-13$ magnitude is generally not a
problem of a sufficient photon rate, even for telescopes as small as
$\sim 20''$.  Rather the difficulties are due to the limited dynamic
range, the scarcity of comparison stars of similar magnitude, and systematic 
errors in the photometry, 
such as correlated noise (Pont et al.\ 2004).  In
this sense, large-format, high-quality, monolithic cameras on relatively small
aperture telescopes are ideal.  As stated, the most impressive results
on follow-up of bright transiting planets have come from the Transit
Light Curve (TLC) project using KeplerCam.
This camera has a single 4K $\times$ 4K Fairchild 486 CCD with a $23'
\times 23'$ field of view.  Using 2 $\times$ 2 binning, the readout
time for this detector is 11.5 s.  The TLC project has demonstrated the
ability to achieve $\sim 0.15\%$ photometry in the $z$-band (where
limb darkening is minimized), with 30 second exposures on the bright
$(V=11.79)$ transit host star TrES-1 (Winn et al.\ 2006).  They furthermore
demonstrated that the uncertainties were essentially uncorrelated,
such that by binning on $\sim 30$ minute timescales, they achieved RMS
residuals of $\sim 10^{-4}$.  The times of individual transits were
measured to a precision of $\sim 10$ seconds.

Currently, there are 9 stars with transiting planets that are sufficiently bright
for precise transit timing measurements.  In principle
all of these could be searched for transit timing signals using
existing observatories.  In practice, there are several barriers.
While transit timing can be performed with a relatively modest aperture,
it does require a high-quality, large-format camera.  Relatively few small
observatories are outfitted with such a detector. Second, the
observations must occur at a specific time, so scheduling can be
difficult, particularly for observatories that scheduled by the night
and/or are shared by multiple institutions.  Third, the transit of a
typical hot Jupiter lasts less than two hours.  This also makes it
difficult to effectively use an observatory with a per-night
scheduling system.  Finally, weather and day-night aliasing means that
it is difficult to observe many consecutive transits from a single site. 
These difficulties will be exacerbated when ongoing transit search programs 
produce more transiting planets in the near future.  
Clearly, a proper TTV study of a significant fraction of the known transiting planets
over the next decade will require additional
observatories.

The observational needs of TTV observations of current and
future systems could best be met by a network of longitudinally
distributed telescopes that can monitor the skies continuously.  These
telescopes need not be particularly large and could be fully
automated.  The entire network could be as small as four telescopes
and as large or larger than 10 (depending upon the needs and the
number of discovered transiting systems).  Here we present a
strawman proposal for such a network, consisting of a number of 20''
dedicated telescopes.  This network would accomplish the majority of the 
science goals outlined above, and at a low cost relative
to other major exoplanet initiatives.  

Each telescope, which can be bought off-the-shelf from a number of
manufacturers, can be equipped with a large-format camera (also
off-the-shelf).  To this end, we consider a 20'' roboticized telescope
with a 2K $\times$ 2K $z$-band optimized detector with a $\sim
26'\times 26'$ FOV.  This setup can achieve a photon collection rate
that is within a factor of $\sim 2$ of the FLWO 1.2m telescope in the
$z$-band.  A filter that is more broad may achieve a higher photon rate.  
Scaling from the results of (Winn et al.\ 2006), this
setup should
achieve $\sim 1.5~{\rm mmag}$ RMS photometry in the $z$-band in one
minute samples on a $V \sim 12$ star.  As stated above, this will also
allow to measure transit times to tens of seconds (Winn et al.\ 2006), and so 
sensitivity to resonant Earth-mass planets.  Based on actual price
quotes, we estimate the total cost for hardware for each such
telescope to be roughly \$100k, along with additional costs for
site fees and computing resources.  For a six-year science program the
total cost of the equipment, its operation, and maintenance is
approximately \$250k per telescope.

A typical planet on a 3-day orbit is in transit for roughly 3\% of its
orbit.  Thus, of the $\sim 10$ known planets there is one in transit nearly
1/3 of the time.  If we assume that a planetary system is only visible
during six months then once $\sim 70$ systems are discovered we expect a
planet to be in transit at all times.  If the Kepler and CoRoT
missions identify even a moderate fraction of their anticipated
yield, then additional telescopes (perhaps located in pairs to mitigate
against coincidence losses or
separated by a few hundred miles to mitigate against weather) would be required to search most of
the expected discoveries of stars hosting transiting planets.  A
network of small telescopes could easily and rapidly be expanded and optimized to
meet the increasing demands of new discoveries.  This capability
represents a significant advantage for the choice of small, low-cost
telescopes.

An even larger potential resource for ground-based transit timing studies
is a network of 0.4 meter and 1 meter telescopes is being built with
funding from Wayne Rosing to study extragalactic transients and
extrasolar planets, with microlensing and transits, called
Las Cumbres Observatory Global Network (LCOGT, Brown et al. 2006).
By the end of 2007 the first of 30 0.4 meter telescopes will be 
commissioned, while by the end of 2008 the first of 30 1-meter 
telescopes will be commissioned which will utilize the same CCDs as 
KeplerCam with a 4-second readout.
Thus, the capability of each of these 60 telescopes will be 
very similar to the strawman telescope proposal described here, while it 
is expected that about 15-20\% of the LCOGT observing time will be devoted 
to planetary transit observations.  These telescopes will be 
distributed longitudinally which is ideal for complete transit
lightcurves. In addition to these telescopes, there are currently
two two-meter robotically controlled telescopes (one in Hawaii and
one in Australia) with fast readouts that are part of the LCOGT network. 
These telescopes are already in operation and will, in principle, allow high precision 
photometry of fainter transiting planet systems (Tim Brown, priv. comm.).

{\bf Transit Timing Analysis:}

Inferring the orbital elements and mass of a perturbing planet via
TTV is generally more complex than other
planet detection schemes.  In transit timing data, the
signal is a combination of several effects including the reflex motion
of the star, the mutual gravitational interaction between the planets,
the changing light travel time, or changing tidal field of a distant
companion (e.g., Borkovits et al.\ 2003; Heyl \& Gladman 2006), though
typically only one effect dominates.
Unlike other dynamical detection techniques, the salient
characteristic of the TTV approach is the deviations from Keplerian
orbits.  This requires high accuracy n-body simulations of each model
in order to calculate its TTV signature (Holman \& Murray 2005; Agol
et al.\ 2005).  Given the computational requirements of n-body
integrations, practical algorithms must explore a high-dimensional
parameter space efficiently.  While challenging,
preliminary tests involving simulated data show that a large
fraction of systems can be correctly identified with appropriate
analysis techniques (Steffen \& Agol 2006) though much more development of
these techniques remains.

\begin{figure}[ht*]
\epsscale{1.0}
\plottwo{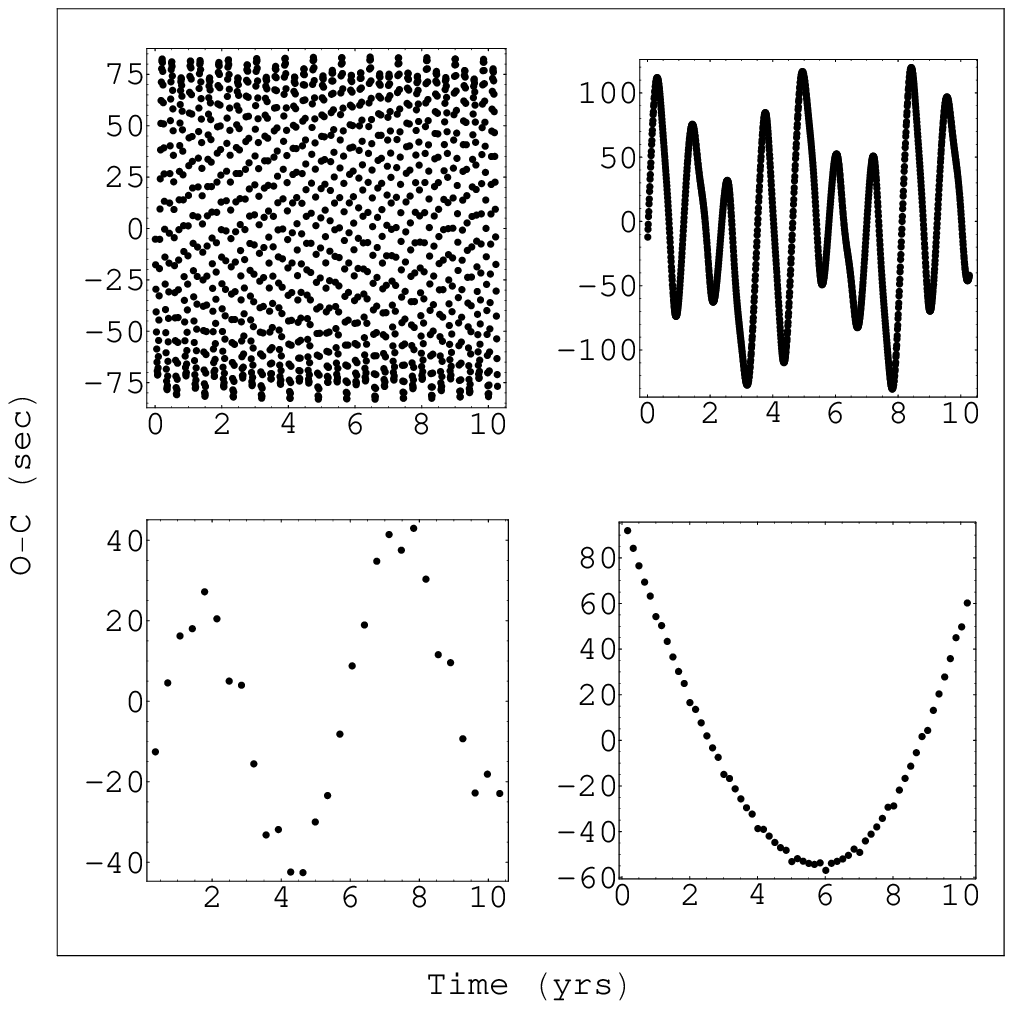}{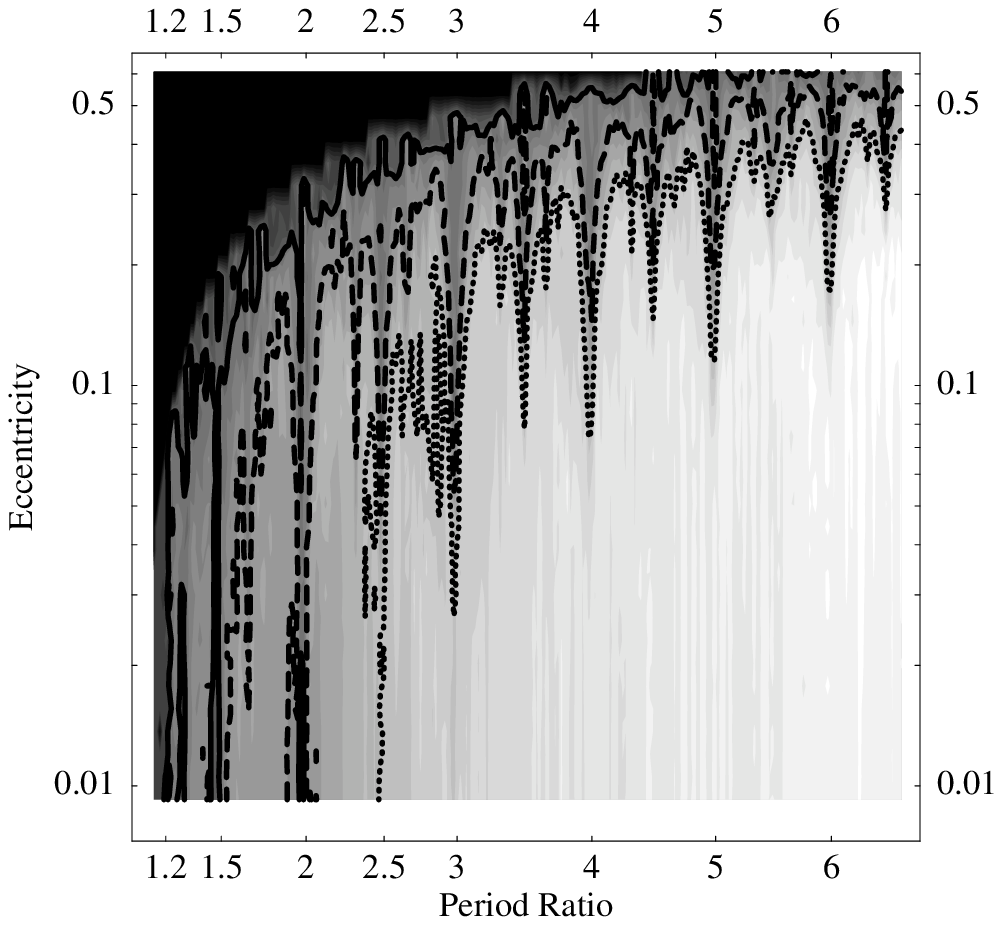}
\caption{
(Left) TTV signal of a Jupiter-mass transiting planet that orbits a solar mass star and is perturbed by an Earth-mass planet where: both planets are on initially circular orbits, the gas giant is on a 3-day orbit, and the perturber is in the interior 2:1 MMR (upper left); the same scenario except the perturber is in the exterior 2:1 MMR (upper right); the gas giant has a period of 130 days, both planets have an eccentricity of 0.05, and the perturber orbits at 1AU---{\it not in MMR} (lower left); the gas giant has an eccentricity of 0.15, the perturber has an eccentricity of 0.25 and orbits in the exterior 6:1 MMR (at 1AU).  (Right) Constraints on the mass of an exterior, secondary planet in the HD209458 system as a function of semi-major axis ratio and eccentricity of the second planet.  We assume that the known planet has an initial eccentricity of zero.  The contours correspond to 100 (dotted), 10 (dashed), and 1 (solid) earth-masses.  The dark region in the upper-left portion of the graph is where the orbits overlap (from Agol \& Steffen 2007).
\label{fig:mt}
}
\end{figure}

{\bf 3. Conclusions}

High precision measurements of the transit times of known of transiting planets provide a method to search for additional low-mass planets in these systems, with sensitivity to (sub) Earth-mass planets and planets near mean-motion resonances.  Thus a serious transit timing campaign would provide constraints for planet formation theory, including migration history, tidal interactions, and the frequency of low-mass planets in resonances.  We expect that the telescopes currently making transit timing observations will soon be overwhelmed by detections from both ground and space-based transit searches.  A network of small, inexpensive telescopes could rapidly be deployed to search for Earth-mass planets around solar-type stars and to address many outstanding scientific questions about planet formation.  Such a network would require a relatively modest investment, and would be relatively easy to scale-up as additional transit searches discover more transiting planets.


\begin{multicols}{2}{
\footnotesize
{\bf References:}\\
Agol E.\ et al., 2005, MNRAS, 359, 567\\
Agol E., \& Steffen, J. 2007, MNRAS, 374, 941\\
Alonso R., et al. 2004, ApJ, 613, L153\\
Baglin A., \& COROT Team 2002, ESA SP-485: Stellar Structure \& Habitable Planet Finding, 17\\
Bakos G.\ A., et al. 2006, ApJ, 656, 552\\
Bodenheimer P.\ et al., 2001, ApJ, 548, 466\\
Borkovits T.\ et al., 2003, A\&A, 398, 1091\\
Borucki W., et al.\ 2003, Proc. SPIE, 4854, 129\\
Brown T.\ M., et al.\ 2001, ApJ, 552, 699\\
Brown, T. et al., 2006, BAAS 208, \#56.05\\
Charbonneau D. et al.\ 2007, PPV, 701\\
Chatterjee S.\ et al., 2007, astro-ph/0703166\\
Collier Cameron A., et al. 2007, MNRAS, accepted\\
Cresswell P.\ \& Nelson, R.\ P., 2006, A\&A, 450, 833\\
Fogg M.\ J., \& Nelson, R. P. 2007, A\&A, 461, 1195\\
Heyl J.\ \& Gladman, B., 2006, astro-ph/0610267\\
Ford E.B., \& Gaudi, B.S. 2006, ApJ, 652, L137\\
Holman M.\ \& Murray, N.\ 2005, Science, 307, 1288\\
Holman M.\ J.\ et al., 2006, ApJ, 652, 1715\\
Juric M.\ \& Tremaine, S., 2007, astro-ph/0703160\\
Knutson H.\ et al.\, 2007, Nature, in press\\
Mandell A.\ M. et al., 2007, astro-ph/0701048\\
McCullough P.\ R., et al. 2006, ApJ, 648, 1228\\
Miralda-Escud\'e J. 2002, ApJ, 564, 1019\\
Narayan R.\ et al., 2005, ApJ, 620, 1002\\
O'Donovan F.\ T., et al. 2006, ApJ, 651, L61 713, 185\\
Pont F.\ et al.\ 2006, MNRAS, 373, 231\\
Steffen J. H., \& Agol, E. 2005, MNRAS, 364, L96\\
Steffen J. H., \& Agol, E. 2007, astro-ph/0612442\\
Terquem C.\ \& Papaloizou, J., 2007, ApJ, 654, 1110\\
Thommes E. W. 2005, ApJ, 626, 1033\\
Winn J.N.\ et al., 2006, ApJ, accepted\\
Wright J.\ T.\ 2005, PASP, 117, 657\\
Zhou J.\ et al., 2005, ApJ, 631, L85}
\end{multicols}
\end{document}